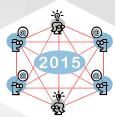

# Skip This Paper

## RINASim: Your Recursive InterNetwork Architecture Simulator

Vladimír Veselý, Marcel Marek, Tomáš Hykel, Ondřej Ryšavý
Department of Information Systems, Faculty of Information Technology
Brno University of Technology
Brno, Czech Republic
{ivesely, imarek, rysavy}@fit.vutbr.cz; xhykel01@stud.fit.vutbr.cz

*Abstract*—Recursive InterNetwork Architecture is a clean-slate approach to how to deal with the current issues of the Internet based on the traditional TCP/IP networking stack. Instead of using a fixed number of layers with dedicated functionality, RINA proposes a single generic layer with programmable functionality that may be recursively stacked. We introduce a brand new framework for modeling and simulation of RINA that is intended for OMNeT++.

*Keywords—Recursive InterNetwork Architecture; RINA; OMNeT++; Delta-t;*

## I. INTRODUCTION

Thank you for not skipping this paper and now straight to the point. RFC 4984 [1] and a follow-up RFC 6227 [2] mention some of the current issues of the Internet. Cumbersome mobility and multihoming (mostly due to the fact of incomplete naming scheme where IP address identifies an interface, not the device), missing native Quality of Service (QoS) support, prefix deaggregation and overall unscalable growth of the default-free zone (backbone of the Internet) are among these problems. Since 2006, many alternatives have been extensively discussed and investigated (e.g., Locator/ID Split Protocol, Identifier-Locator Network Protocol, Multipath TCP, etc.). However, none of them were able to come up with a unifying solution to all problems.

**Recursive InterNetwork Architecture (RINA)** is a complete replacement of the traditional TCP/IP (or ISO/OSI) layered network model based on a rigid differentiation between functionalities of each layer. RINA (initially proposed by John Day's book [3]) is a subject of ongoing research interest and initiatives during last few years. RINA's core principle says that *networking* is nothing more than only inter-process communication (IPC). The IPC happens over different scopes, where the layer limits a scope. Layers may be recursively stacked to provide IPC services to applications and/or layers above. Layer in RINA is called **Distributed IPC Facility (DIF)**, and it is formed by a set of cooperating IPC processes. Each DIF is completely independent, offering wide gamut of mechanisms in order to achieve: (un)reliable data transfer; flow and congestion control; relaying and multiplexing of traffic; routing decisions or other DIF management tasks; etc. RINA clearly distinguishes between mechanisms and policies, which means that all DIFs offer the same set of mechanisms but operate under potentially different configurations (policies).

The purpose of this paper is to present the community with RINASim – the RINA-capable framework for the OMNeT++ simulator – and describe its components and basic order of operations. This paper has the following structure. The next section covers a brief introduction to some of RINA concepts. Section III reveals components of our framework together with their description. Section IV describes flow lifecycle in two simple scenarios where a pair of hosts is exchanging data. The paper is summarized in Section V together with unveiling of our plans.

## II. STATE OF THE ART

This section introduces basic RINA principles and terminology. However, astute reader is advised to consult additional information sources (see [4], [5] or [6]) because not everything can be explained due to the limited space of this paper.

Nature of *applications* in RINA is as follows: **Application Process (AP)** is program instantiation to accomplish some purpose; **Application Entity (AE)** is the part of an AP, which represents the application protocol and application aspects concerned with the IPC. There may be multiple instances of the Application Process in the same system. AP may have multiple AEs, each one may process a different application protocol. There also may be more than one instance of each AE type within a single AP. All application protocols are stateless; the state is and should be maintained in the application. Thus, all **application protocols** modify shared state external to the protocol itself on various objects (e.g., data, files, HW peripherals). Because of that, there is only one application protocol that contains trivial operations (e.g., read/write, start/stop, and create/delete).

As it was mentioned before, RINA separates mechanisms (fixed parts of a system) from policies (variable parts of a system) of any IPC. There are two types of mechanisms: a) **tightly-bound** that must be associated with every packet, which handle fundamental aspects of data transfers; and b) **loosely-bound** that may be associated with packet and which provide additional features (namely reliability and flow control). Both types are coupled through a state vector maintaining state information. Previous implies the existence of a single soft state **data transfer protocol**. This protocol controls data transfer with the help of different policies. Initial synchronization of communicating parties is done with the help of **Delta-t** protocol see ([7], and [8]). Delta-t was developed by Richard Watson, who proposed a timer-based synchronization technique. He proved that conditions for distributed synchronization are met if following three timers are bound: a) Maximum Packet Lifetime (MPL); b) Maximum time to attempt retransmission a.k.a. maximum period during which a sender is holding a packet for retransmission when waiting for a positive acknowledgement ($R_{timer}$); c) Maximum time before Acknowledgement ($A_{timer}$). Delta-t assumes that all connections exist all the time. The synchronization state is maintained only during the data transfer activity, but after $2\Delta t$ or $3\Delta t$ periods without any traffic the state may be discarded which effectively resets the connection, where $\Delta t = MPL + A_{timer} + R_{timer}$. Because of that, there is no need for hard state (with explicit synchronization using SYNs and FINs). Delta-t postulates that port allocation and synchronization are distinct.

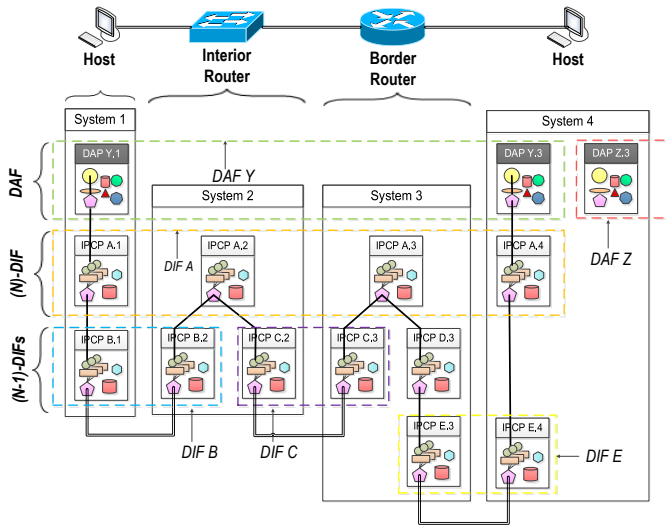

Fig. 1. DIF, DAF, DAP and IPCP illustration

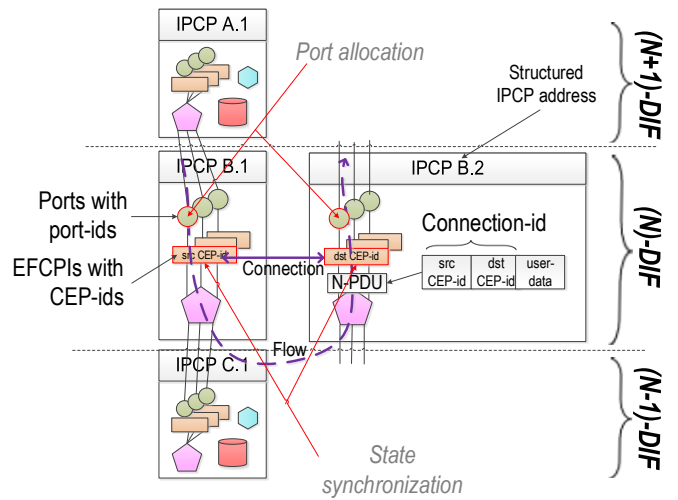

Fig. 2. IPCP's local identifiers overview

The concept of RINA layer a.k.a. DIF could be further generalized to **Distributed Application Facility (DAF)** – a set of cooperating APs in one or more computing systems, which exchange information using IPC and maintain shared state. A DIF is a DAF that does only IPC. **Distributed Application Process (DAP)** is a member of a DAF. **IPC Process (IPCP)** is a special AP within a DIF delivering inter-process communication. IPCP is an instantiation of a DIF membership; computing system can perform IPC with the other DIF members via its IPC process within this DIF. An IPCP is specialized DAP. The relationship between all newly defined terms is depicted in Fig. 1.

We need to differentiate between different APs and also different AEs within the same AP. Thus, RINA uses **Application Process Name (APN)** as a globally unambiguous, location-independent, system-dependent name. **Distributed Application Name (DAN)** is a globally unambiguous name for a set of system-independent APs. An IPC Process has an APN so it can be identified among other DIF members. An **address** is a synonym for the IPCP's APN with a scope limited to the layer and structured to facilitate forwarding. An APN is useful for management purposes but not for forwarding. The address structure may be topology-dependent (indicating the nearness of IPCPs). An APN and an address are simply two different means to locate an object in different context. There are two local identifiers important for the IPCP functionality – a port-id and a connection-endpoint-id. **Port-id** binds this (N)-IPCP and (N+1)-IPCP/AP; both of them uses the same port-id when passing messages. Port-id is returned as a handle to the allocator and is unambiguous within a computing system. **Connection-endpoint-id (CEP-id)** identifies a shared state of one communication endpoint. Since there may be more than one flow between the same IPCP pair, it is necessary to distinguish them. For this purpose, **Connection-id** is formed by combining source and destination CEP-ids with QoS requirements descriptor. CEP-id is unambiguous within an IPCP and the Connection-id is unambiguous between a given pair of IPCPs. Fig. 2 depicts all relevant identifiers between two IPCPs. Watson's Delta-t implies Port-id and CEP-id to help separate port allocation and synchronization. The RINA **connection** is a shared state between protocol machines – ends identified by CEP-ids. The RINA **flow** is bound to ports identified by port-ids. The lifetimes of a flow and its connection(s) are independent of each other.

III. CONTRIBUTION

We are developing **RINASim** in the frame of FP7 project PRISTINE [9]. The goal is to provide the public community with a full-fledged RINA simulator to support ongoing research and academic activities. To understand RINA architecture means to understand each of its elements. This subsection starts with a description of high-level RINA network nodes and then goes deeper and outlines various implemented components.

There are only three basic kinds of nodes in RINA network. Each kind represents computing system running RINA: a) **hosts**, IPC end-devices containing AEs, they employ two or more DIF levels; b) **interior routers**, interim devices interconnecting (N)-DIF neighbors via multiple (N-1)-DIFs, they employ two or more DIF levels; c) **border routers**, interim devices interconnecting (N)-DIF neighbors via (N-1)-DIFs, where some of (N-1)-DIFs are reachable only through (N-2)-DIFs, they employ three or more DIF levels. Fig. 1 depicts simple RINA network containing all kinds of nodes and their basic internal structure.

The internal structure of any RINA node could be divided into two parts – the one responsible for DAF operation and another one for DIF operation. DAF part contains one or more APs and a couple of management components, which are described in Table I. DIF part contains one or more IPCPs of different ranks interconnected to create a recursive stack. All IPCPs consist of a same set of subcomponents, which are depicted in Fig. 3 and summarized in Table II.

TABLE I. DAF COMPONENTS OVERVIEW

| Name | Description |
|---|---|
| application Process | AP module contains one or more AE instances. AE processes application protocol and uses IPCP to communicate with destination AE(s). |
| ipcResource Manager | IPC Resource Manager (IRM) manages DAF resources, which involve delegation of flow (de)allocation calls or management of a new DAF/DIF join or creation. IRM maintains information about all connections used by AE(s). |
| dif Allocator | DIF Allocator (DA) primary task is to return a list of DIFs where destination application may be found given APN and access control information. DA contains and works with multiple mapping tables to provide its services. |

RINA's generic approach to an IPCP's operation is met via NED interfaces. Each policy is represented as a simple module implementing specific interface. NED interfaces enabled us to use different implementations for a specific policy, which is not possible to do directly in C++. Policies related to static modules (RMT, FA) are specified via variables in a simulation configuration file (.ini) and policies related to dynamic modules (EFCP instances) are configured by an XML configuration file. Dynamic modules cannot be addressed in the .ini file before the start of the simulation because their names are not yet known due to randomness in generated names.

All inner IPCP interconnections are modeled with zero-time delay since processing messages between modules is a matter of a queue scheduling algorithm and not a network architecture. The bottommost interconnection between IPCPs represents the physical medium offering to set various properties (rate, delay, bit error rate). The recursiveness enables us to set the properties on all interconnections between IPCPs.

TABLE II. IPCP COMPONENTS OVERVIEW

| Name | Description |
|---|---|
| flow Allocator | Flow Allocator (FA) processes (de)allocate calls. FA creates a Flow Allocator Instance (FAI), which manages each flow independently. FAI spawns separate EFCP instance to handle data and interconnects all involved components with bindings to create data-path. FA translates application QoS requirements onto available RA's QoS profiles. |
| efcp | Error and Flow Control Protocol (EFCP) provides data transfer services, and it is split into two independent parts. Data Transfer Protocol (DTP) implements mechanisms tightly coupled with transported SDUs, e.g., fragmentation, reassembly, and sequencing. DTP works with packet header fields like source/destination addresses, QoS requirements, Connection-id, optionally sequence number or checksum. Data Transfer Control Protocol (DTCP) implements mechanisms that are loosely coupled with transported SDUs, e.g., (re)transmission control using various acknowledgment schemes and flow control with data-rate limiting. DTCP functionality is based on Delta-t. |
| relay And Mux | Relay and Multiplexing (RMT) instances in hosts and bottom layers of routers usually perform just the multiplexing task, while RMTs in top layers of interior/border routers do both multiplexing and relaying. RMTs in top layers of border routers perform aggregation of traffic from multiple (N)-flows onto single (N-1)-flow. Each (N-1)-port handled by RMT has own set of input and output buffers. RMT performs routing decision putting PDUs into proper queues of (N-1)-flows leading to various destinations. |
| resource Allocator | If a DIF has to support diverse QoS, then different flows will have to be allocated to different policies and traffic for them treated differently. Resource Allocator (RA) is component accomplishing this goal by monitoring and managing RMT queues, EFCP policies, available QoS profiles, etc. |
| ribDaemon | All management information maintained by IPC components such as FA, RA, etc. is available and updated through RIB Daemon (RIBd). RIBd is sending and receiving management message on behalf of other components. |
| routing Policy | This module is pure policy computing an optimal path to other destinations by given metrics. Usually some routing algorithm exchanges information with other IPCP members of a DIF. |
| enroll ment Module | Enrollment module is involved when IPCP is joining a DIF. At this time, Enrollment module finite-state machine governs the exchange of authentication information during the connection establishment phase. |

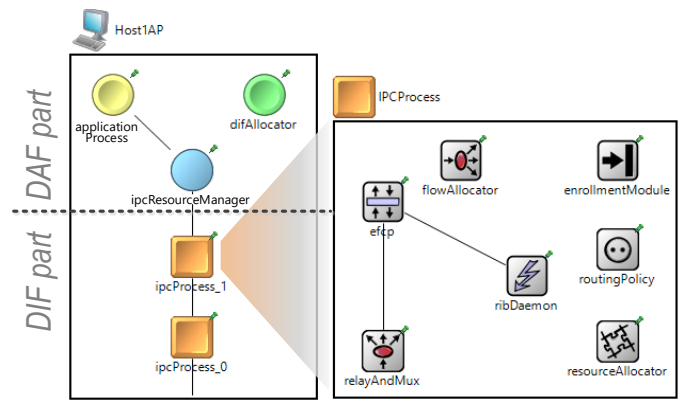

Fig. 3. Host node and IPC Process structure

## IV. TESTING

In this section, we present example scenario to illustrate the flow lifecycle. Simulation topology (shown in Fig. 4) consists of two hosts (`Host1` and `Host2`) and one interior router (`SW`) performing relaying between the two hosts. Hosts APs `AP_A` and `AP_B` employ a ping-like application protocol exchanging request-response messages and measure one-way/round-trip time latencies.

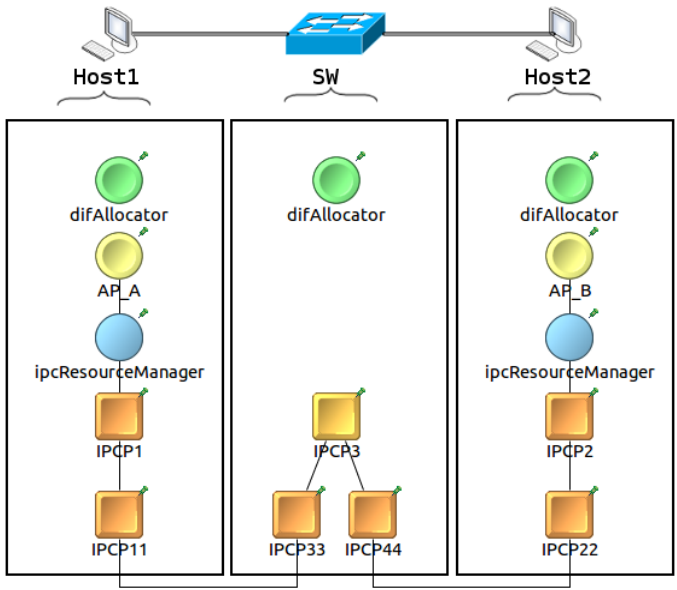

Fig. 4. Testing scenario

The flow initialization can be separated into five phases. The odd phases belong to `Host1`, and even phases belong to `Host2`. The first phase describes beginning of the allocation.

1) `AP_A` on `Host1` notifies IRM with its request;
2) IRM notifies `IPCP1`'s FA to check whether there is any flow available between applications `A` and `B`. Since there is neither data nor management (N)-flow, `IPCP1`'s FA invokes management flow allocation in order to create IPC channel for exchange of management messages;
3) `IPCP1`'s FA asks `IPCP1`'s RA to prepare (N-1)-flow for management. Flow allocation process is recursively repeated from the beginning whenever there is no available flow in the underlying DIF until it reaches medium, where flow allocation is considered inherent.

During the second phase, `IPCP2`'s FA of `Host2` receives allocation request, which is accepted. Subsequently, bindings between bottom `IPCP22` and `IPCP2`'s RIBd are formed.

During the third phase, `Host1` `IPCP1`'s FA receives a positive allocation response for management flow and starts enrollment procedure. During enrollment `Host1` and `Host2` exchanges authentication data. Upon successful enrollment, `Host1` continues initial data flow allocation.

1) `Host1`'s FA creates FAI for data flow. FAI spawns EFCP instance. Following next, FAI prepares bindings and lets RIBd send *CreateFlowRequest*;
2) *CreateFlowRequest* is processed by `IPCP3`'s FA where it needs to be forwarded to `IPCP44`. But before that underlying (N-1)-flow are recursively allocated, which is secured by `IPCP3`'s RA.

During the fourth phase, `Host2` handles *CreateFlowRequest* for data flow.

1) `IPCP2`'s RIBd notifies `Host2`'s AP about flow allocation and connection attempt. `AP_B` accepts and governs IRM to prepare bindings and delegate flow allocation request;
2) IRM invokes `IPCP2`'s FA, which instantiates FAI. FAI spawns EFCP instance to provide data transfer service to the data flow. FAI prepares bindings and lets RIBd send positive *CreateFlowResponse*.

During the fifth phase, `IPCP1`'s RIBd receives a positive *CreateFlowResponse* and notifies about it FAI. Then, FAI notifies the application about the successful finalization of the flow allocation. Subsequently, the actual ping-like data transfer occurs between and using allocated data-paths.

When APs finish data exchange, the flow deallocation process is initiated. Deallocation is separated into three phases.

During the first phase, `AP_A` initiates data flow deallocation.

1) Before actual deallocation, `AP_A` releases connection notifying `AP_B` about this fact;
2) Following next, `AP_A` tells IRM about deallocation request, which delegates it to appropriate `IPCP1`'s FAI;
3) `IPCP1`'s FAI asks RIBd to deliver *DeleteFlowRequest*.

During the second phase, `AP_B` receives dellocation request.

1) `IPCP2`'s FAI on `Host2` receives *DeleteFlowRequest*. It notifies `AP_B` about it, and initiates data flow deallocation by removing appropriate EFCP instance and disconnecting bindings;
2) RIBd of `IPCP2` acknowledges deallocation by sending to `IPCP1` *DeleteFlowResponse* on behalf of FAI. At this point, data flow and data-path from `AP_B` side on `Host2` are deallocated.

During the third phase, deallocation is finalized on `AP_A` side. `IPCP1`'s FAI receives a *DeleteFlowResponse* and finishes data flow deallocation by removing EFCP and relevant bindings. Deallocation is complete, and port-IDs associated with `AP_A` and `AP_B` may be reused by other APs.

Simulation results and message confluence have been successfully verified against proposed behavior in RINA specification.

V. CONCLUSION

We have outlined the basic principles behind RINA, a clean-slate replacement of the traditional TCP/IP stack. We have introduced RINASim as an independent OMNeT++ framework providing simulation environment for educational and research purposes with this fresh architecture. We plan to carry on work and further refine our framework based on new knowledge and up-to-date specifications. An additional goal is to conduct a comparative evaluation of our simulation models with RINA implementation for Linux environment called IRATI [10]. All source codes are publicly available on GitHub repository [11] under the MIT license. We encourage the reader to read accompanied documentation [12] or generate Doxygen documentation to get more insight.

ACKNOWLEDGMENT

This work was supported by the Brno University of Technology organization and by the research grants: FP7-PRISTINE supported by European Union; FIT-S-14-2299 supported by Brno University of Technology; IT4Innovation ED1.1.00/02.0070 supported by Czech Ministry of Education Youth and Sports.

REFERENCES

[1] D. Meyer, L. Zhang, and K. Fall. (2007, September) RFC 4984: Report from the IAB Workshop on Routing and Addressing. [Online]. http://tools.ietf.org/html/rfc4984

[2] T. Li. (2011, May) RFC 6227: Design Goals for Scalable Internet Routing. [Online]. http://tools.ietf.org/html/rfc6227

[3] J. Day, *Patterns in Network Architecture: A Return to Fundamentals*, 1st ed. ISBN-13: 978-0137063383: Prentice Hall, 2008.

[4] J. Day, I. Matta, and K. Mattar, "Networking is IPC: a guiding principle to a better internet," in *Proceedings of the 2008 ACM CoNEXT Conference*, NY, USA, 2008, p. 6.

[5] E. Trouva, E. Grasa, J. Day, and S. Bunch, "Layer discovery in RINA networks," in *IEEE 17th International Workshop on Computer Aided Modeling and Design of Communication Links and Networks (CAMAD)*, Barcelona, Spain, 2012, pp. 368 - 372.

[6] E. Trouva et al., "Is the Internet an unfinished demo? Meet RINA!," in *TERENA Networking Conference*, Prague, Czech Republic, 2011, p. 12.

[7] R. Watson, "Delta-t protocol specification," Lawrence Livermore National Lab., California, USA, UCID-19293, 1981.

[8] R. Watson, "The Delta-t transport protocol: features and experience," in *Proceedings 14th Conference on Local Computer Networks*, Mineapolis, MN, USA, 1989, pp. 399-407.

[9] PRISTINE consortium. (2015) PRISTINE | PRISTINE will take a major step forward in the integration of networking and distributed computing. [Online]. http://ict-pristine.eu/

[10] F. Salvestrini. (2015) IRATI · GitHub. [Online]. https://github.com/IRATI

[11] Brno University of Technology. (2015) kvetak/RINA · GitHub. [Online]. https://github.com/kvetak/RINA

[12] V. Veselý. (2015) RINA Simulator: basic functionality. [Online]. http://ict-pristine.eu/wp-content/uploads/2013/12/PRISTINE-D24-RINASim-draft.pdf